\documentclass[
    aps,
    prb,
    reprint,
    superscriptaddress,
    showkeys,
    showpacs,
    amsmath,
    amssymb,
    floatfix,
    citeautoscript,
    ]{revtex4-2}
\usepackage[shortlabels]{enumitem}
\usepackage{multirow}
\usepackage{booktabs}
\usepackage{tabularx}
\usepackage{bm}
\usepackage{graphicx}
\usepackage{subcaption}
\usepackage{xspace}
\usepackage[
    colorlinks=true,
    linkcolor=blue,
    citecolor=blue,
    urlcolor=blue,
    ]{hyperref}
\usepackage[capitalise]{cleveref}
\usepackage{url}
\usepackage{orcidlink}
\newcommand{\QE}{\textsc{Quantum ESPRESSO} (QE)\xspace}

\begin{document}
\title{BMach: a Bayesian machine for optimizing Hubbard U parameters in DFT+U with machine learning}
\author{Ritwik Das}
\email[]{ritwik.das@universite-paris-saclay.fr}
\homepage[\newline]{https://www.c2n.universite-paris-saclay.fr/en/}
\affiliation{Centre for Nanoscience and Nanotechnology (C2N) --- French National Centre for Scientific Research (CNRS), Paris, France}
\affiliation{Université Paris-Saclay, Paris, France}

\begin{abstract}
    Accurately determining the effective Hubbard parameter ($U_{eff}$) in Density Functional Theory plus U (DFT+U) remains a significant challenge, often relying on empirical methods or linear response theory, which frequently fail to predict accurate material properties. This study introduces BMach, an advanced Bayesian optimization algorithm that refines $U_{eff}$ by incorporating electronic properties, such as band gaps and eigenvalues, alongside structural properties like lattice parameters. Implemented for the \textsc{Quantum ESPRESSO} distribution, BMach demonstrates superior accuracy and reduced computational cost compared to traditional methods. The BMach-optimized $U_{eff}$ values yield electronic properties that align closely with experimental and high-level theoretical results, providing a robust framework for high-throughput materials discovery and detailed electronic property characterization across diverse material systems.
\end{abstract}

\maketitle

\section[short]{INTRODUCTION}\label{sec:intro}

Density Functional Theory (DFT) has emerged as a cornerstone in computational materials science, providing invaluable insights into the electronic structures of various materials.
DFT's widespread adoption can be attributed to its impressive balance between computational efficiency and the accuracy of predictions.
However, when applying standard Kohn-Sham (KS) DFT, which utilizes local or semilocal approximations for exchange ($E_x$) and correlation ($E_c$) energies, certain limitations emerge.
These include the lack of integer discontinuity in the exchange-correlation energy ($E_{xc}$), potential discontinuities when altering electron numbers, and self-interaction errors (SIE).
Such limitations lead to an underestimation of the KS one-electron band gaps compared to the experimental and quasiparticle (QP) band gaps, occasionally to the extent that erroneously predicts semiconductors to be metallic \cite{jones_density_1989, mori-sanchez_localization_2008}.
These distinctions are profound in semiconductor application, and the predictive accuracy in band topologies and effective masses is a mere academic concern and quintessential for device engineering \cite{kittel_introduction_2018}.
For high-throughput materials discovery efforts, the widely used conventional semi-local exchange-correlation functionals, like the generalized gradient approximation (GGA) of Perdew, Burke, and Ernzerhof (PBE) \cite{perdew_generalized_1997}, fall short in addressing the SIE issues.
There are two distinct approaches to mitigating the nuances mentioned above and obtaining a comprehensive description of electronic band characteristics.
The first approach involves augmenting a portion of the fully nonlocal exact Hartree-Fock (HF) exchange in hybrid exchange-correlation (XC) functionals, such as done in the Heyd-Scuseria-Ernzerhof (HSE) hybrid XC functional \cite{heyd_erratum_2006}.
The second approach employs a quasiparticle Green function-based GW approximation, rooted in many-body perturbation theory, and GW coupled with the Bethe-Salpeter equation (GW-BSE) approach, offering an effective method for analyzing electronic band properties \cite{zhu_quasiparticle_1991}.
Both of these methods are well-regarded for their ability to yield enhanced band structures and band gaps.
Nevertheless, their significant computational demands can be a limiting factor, particularly for expansive systems.
This includes interface models composed of several hundred atoms and scenarios where a vast array of materials requires screening.

Another significant stride in addressing this limitation is the DFT+U method, initially proposed by Anisimov et al. (1991) \cite{anisimov_band_1991} and subsequently refined by Liechtenstein et al. (1995) \cite{liechtenstein_density-functional_1995} and Dudarev et al. (1998) \cite{dudarev_electron-energy-loss_1998}.
This DFT+U methodology employs a Hubbard-like model to rectify self-interaction (SIE) errors with the system's total energy delineated by \cref{eq:hubbardeq}.

\begin{equation}
    E_{tot} = E_{DFT} + \frac{U - J}{2} \sum_{\sigma} n_{m,\sigma} - n_{m,\sigma}^2
    \label{eq:hubbardeq}
\end{equation}

where $ n $ represents the number of electrons occupying an atomic orbital, $ m $ stands for the orbital momentum, and $ \sigma $ indicates the spin state.
Here, $ U $ denotes the on-site Coulomb repulsion, and $ J $ refers to the exchange interaction between electrons.
To further refine this approach, the exchange interaction $ J $ is often amalgamated with the Coulomb term $ U $, leading to the definition of an effective Hubbard parameter, $ U_{eff} = U - J $.
The efficacy and accuracy of the DFT+U calculations greatly depend on the precise selection of the $ U_{eff} $ value, which is tailored to suit the specific characteristics of the system under investigation.

Despite this advancement, the determination of the Hubbard $ U_{eff} $ parameter remains pivotal yet poses substantial challenges.
Predominantly, $ U_{eff} $ has been ascertained through empirical methods, which involve aligning the parameter with experimental findings, such as accurately replicating material-specific band gaps.
Though beneficial for their practicality, the conventional reliance on empirical or semi-empirical techniques often does not fully encapsulate the breadth of electronic interactions within diverse materials.
This approach also encounters a crucial obstacle when experimental data is unavailable - a frequent scenario in exploring new materials.
Innovative first-principle strategies have been explored to accurately determine the Hubbard U parameter to bridge this gap.
Among these, the linear response (LR) methodology stands out, rooted in constrained DFT (CDFT) \cite{cococcioni_linear_2005}.
This technique imposes a linear relationship between the total energy and the occupation numbers, addressing the unrealistic energy curvature seen in local and semi-local functionals.
Here, $ U_{eff} $ is derived from the disparity between the inverse non-interacting and interacting density responses, symbolized as $ \chi_0^{-1} $ and $ \chi^{-1} $, respectively.
This disparity is linked to the second derivatives of the non-charge-self-consistent DFT energy ($ E $) and the charge-self-consistent DFT energy ($ E_{KS} $) concerning the localized occupation at a site ($ q_I $).
The equation is elegantly articulated in \cref{eq:ueff}.

\begin{equation}
    U_{eff} = \frac{\partial^2 E[\{q_I\}]}{\partial q_I^2} - \frac{\partial^2 E_{KS}[\{q_I\}]}{\partial q_I^2} = (\chi_0^{-1} - \chi^{-1})_{II}
    \label{eq:ueff}
\end{equation}

Implementing this model necessitates the construction of a super-cell, which simulates occupation variations in an infinite crystal structure.
The size of this super-cell is progressively increased until $ U_{eff} $ converges, often incurring significant computational demands.
Building on this, Kulik et al. have introduced an advanced self-consistent iteration of the LR method \cite{kulik_density_2006}.
Parallel to this, there is the unrestricted Hartree-Fock (UHF) approach, which represents another pivotal first-principles strategy introduced by Mosey et al. \cite{mosey_rotationally_2008}.
This method involves UHF calculations on a finite-sized cluster, representing the bulk material, where the U parameters must reach convergence as the cluster size escalates.
The constrained random-phase approximation (cRPA) \cite{miyake_screened_2008, sasioglu_effective_2011} method also offers another valuable perspective, though computationally more intensive.

In this context, Yu et al. (2020) proposed a groundbreaking technique using Bayesian optimization (BO) for the determination of $ U_{eff} $ \cite{yu_machine_2020}.
This method marks a significant advancement, particularly in computational efficiency and band structure prediction.
However, a critical aspect that warrants further exploration is the relationship between the Hubbard U parameter and lattice parameters.
This relationship is pivotal, as the lattice parameters are not merely passive elements of electronic structure changes but actively influence them, particularly in materials with strong electron correlation effects.
For instance, in some transition metal oxides, a change in the electronic state (like a change from an insulating to a metallic state) can be accompanied by a structural phase transition, indicating a coupling between electronic properties and lattice parameters.

Recognizing this gap, our study introduces the Bayesian Machine (BMach), an advanced algorithm that extends the Bayesian optimization framework to incorporate not only band gaps and eigenvalues but also lattice parameters in optimising the Hubbard U parameters.
Further details of the BMach implementation and its novel contributions are elaborated in the following sections.

By integrating the lattice parameter relationship into the optimization process, BMach improves the fidelity of DFT+U calculations and broadens the scope of materials that can be accurately modelled.
This advancement is crucial for materials where the interplay between electronic and lattice properties is essential for understanding their physical properties.

In summary, BMach represents a significant contribution to computational materials science.
It offers a robust framework for characterizing electronic properties and paves the way for developing novel materials with tailored features.
Our approach underscores the importance of considering the lattice parameter in conjunction with the Hubbard U value, providing a more accurate and comprehensive model for material simulations.

\section[short]{MODELS AND COMPUTATIONAL METHODS}\label{sec:models_methods}

\subsection[short]{BMach: Bayesian Machine Learning Model}\label{subsec:bmach}

Bayesian optimization (BO) is a machine learning algorithm that excels in the global optimization of complex, black-box functions.
It begins by hypothesizing the shape of the unknown function and iteratively refines this estimation by sampling points that either show promise or possess high informational content.
This process is guided by a Bayesian statistical model, typically a Gaussian process, which is favoured in BO algorithms due to its ability to quantify uncertainty in predictions.
The algorithm evolves through the use of an acquisition function, which directs future evaluations of the function based on accumulated data.

The rationale behind BMach is anchored on the principle that the accurate replication of electronic properties necessitates a holistic consideration of all factors influenced by the Hubbard U parameter.
This inclusive approach ensures a comprehensive calibration of U, thereby enhancing the accuracy and reliability of DFT+U calculations.

A key factor in accurately modelling materials using DFT+U is understanding how changes in the Hubbard U value affect the equilibrium lattice parameter.
The Hubbard U value, integral in modelling electron-electron interactions within an atom, can lead to variations in the predicted lattice constants as it adjusts the degree of electron correlation in the calculations.
This interaction is expressed in \cref{eq:hubbardU_A}.

\begin{equation}
    a(U) = a_0 + \alpha (U - U_0)
    \label{eq:hubbardU_A}
\end{equation}

Where $ a(U) $ represents the lattice parameter at a given U value, $a_0$ is the lattice parameter at a reference U value $ U_0 $, generally $ a_0 $ could be taken as the experimental lattice parameter value if the data is available.
$ \alpha $ is a material-dependent coefficient.

\subsubsection[short]{\textbf{Objective Function in BMach}}

Effectively utilizing Bayesian optimization hinges on the selection of a suitable objective function.
In this instance, BMach's surrogate objective function $f(\vec{U})$ encompasses not only the band gaps ($E_g$) and eigenvalues ($\epsilon$) but also the lattice parameters, as delineated in \cref{eq:bmachObjFunc}.
\begin{equation}
    f(\vec{U}) = \alpha_1 (E_g^{Ref} - E_g^{DFT+U})^2 + \alpha_2 (\Delta Band)^2 + \alpha_3 (\Delta A)^2
    \label{eq:bmachObjFunc}
\end{equation}

Where $\Delta A$ and $\Delta Band$ are defined by \cref{eq:delA} and \cref{eq:delBand} \cite{huhn_one-hundred-three_2017}, respectively.

\begin{equation}
    \Delta A = a_0 - a(U)
    \label{eq:delA}
\end{equation}

\begin{equation}
    \Delta Band = \sqrt{\frac{1}{N_E} \sum_{i=1}^{N_k} \sum_{j=1}^{N_b} (\epsilon_{Ref}^j [k_i] - \epsilon_{DFT+U}^j [k_i])^2}
    \label{eq:delBand}
\end{equation}

In this optimized $U_{eff}$ value calculation, the $Ref$ can be corresponded to any advanced and accurate calculation such as HSE or GW calculations.
The weights $\alpha_1$, $\alpha_2$, and $\alpha_3$ are crucial as they dictate each term's relative emphasis in the optimization process.
These weights are meticulously chosen to reflect the relative importance of band gap accuracy, band structure fidelity, and lattice parameter congruence and are not necessarily constrained to sum up to unity (1); instead, they are selected based on the relative significance of each term.
For example, if the band gap error is considered more critical than the lattice parameter deviation, $ \alpha_1 $ might be larger than $ \alpha_3 $.
The choice of these weights is informed by a combination of theoretical insights and empirical adjustments, ensuring that the algorithm adequately reflects the importance of each aspect of the material's properties.

\subsubsection[short]{\textbf{Acquisition Functions in BMach}}

\paragraph{\textbf{Upper Confidence Bound (UCB) Acquisition Function}} \leavevmode\\
In the context of BMach, the Gaussian process forms the backbone of our model, emulating the objective function with high precision and providing a measure of uncertainty at each step.
To balance exploration and exploitation in the optimization process, the Upper Confidence Bound (UCB) Acquisition Function is employed, as defined in \cref{eq:ucbFunction}:

\begin{equation}
    \vec{U}_n = \mathop{\arg \min} \limits_{\vec{U}} [\mu (\vec{U}) + \kappa \sigma (\vec{U})]
    \label{eq:ucbFunction}
\end{equation}

This function effectively balances the trade-off between exploring new possibilities and exploiting known information.
The parameter $ \kappa $ regulates this balance, with higher values favoring exploration over exploitation.

\paragraph{\textbf{Custom Acquisition Function}} \leavevmode\\
Within the Bayesian Machine (BMach) algorithm, a significant enhancement is our bespoke \texttt{BMach\_EI} acquisition function.
This function deviates from the standard Gaussian Expected Improvement (EI) approach, typically employed in Bayesian optimization frameworks.
Our adaptation aims to enhance precision and control in the optimization process, which is crucial for the intricate simulations of materials.

The standard Gaussian EI, commonly utilized in Bayesian optimization, is mathematically expressed as:

\begin{equation}
    EI(\vec{x}) = (\mu_{opt} - \mu(\vec{x}) - \xi) \Phi(Z) + \sigma(\vec{x}) \phi(Z)
    \label{eq:gaussianEI}
\end{equation}

Here, $\mu_{opt}$ symbolizes the optimum observed value thus far, $\mu(\vec{x})$ and $\sigma(\vec{x})$ represent the mean and standard deviation of the prediction at $\vec{x}$, respectively.
$\Phi$ and $\phi$ correspond to the standard normal distribution's cumulative and probability density functions.
The parameter $\xi$ plays a pivotal role in modulating the exploration-exploitation balance.

The \texttt{BMach\_EI} function introduces several key modifications to this conventional EI formulation:

\begin{enumerate}[(i)]
    \item \textbf{Explicit Zero Standard Deviation Handling:}
          \texttt{BMach\_EI} is designed to assign an EI of zero when the standard deviation ($\sigma$) is zero, averting potential numerical instabilities and reflecting the model's confidence in its predictions.
          \begin{equation}
              EI(\vec{x}) =
              \begin{cases}
                  0                                                                  & if \sigma(\vec{x}) = 0 \\
                  (\mu_{opt} - \mu(\vec{x}) - \xi) \Phi(Z) + \sigma(\vec{x}) \phi(Z) & otherwise
              \end{cases}
              \label{eq:BMachEI}
          \end{equation}

    \item \textbf{Custom Exploration Parameter ($ \xi $):}
          The \texttt{BMach\_EI} function incorporates an adjustable exploration parameter, $ \xi $, enhancing the flexibility in controlling the exploration-exploitation balance.
          This adaptability is vital for complex material simulations.
          \begin{equation}
              Z = \frac{\mu_{opt} - \mu(\vec{x}) - \xi}{\sigma(\vec{x})}
              \label{eq:custom_exploration}
          \end{equation}
          Adjusting $ \xi $ directly influences the value of $ Z $, thereby impacting the decision-making process between exploring new areas and exploiting known regions.

    \item \textbf{Adaptive Improvement Calculation:}
          Contrasting the static nature of Gaussian EI, \texttt{BMach\_EI} dynamically adapts its improvement calculation, reflecting the current state of the model.
          This adaptive feature ensures responsiveness in the optimization strategy, particularly for materials with complex electronic structures.
          \begin{equation}
              Imp(\vec{x}) = \mu_{opt} - \mu(\vec{x}) - \xi
              \label{eq:adaptive_improvement}
          \end{equation}
          Imp($\vec{x}$) changes dynamically in each iteration, considering the current optimum value and the model's predictions.
\end{enumerate}

By integrating these theoretical enhancements into the acquisition function, the BMach algorithm emerges as a sophisticated and tailored solution for the multifaceted challenges in material simulations.
These novel adaptations underscore our commitment to advancing computational efficiency and accuracy in the pursuit of novel material discoveries.

\subsubsection{BMach Algorithm Flowchart}

\begin{figure}[!htb]
    \centering
    \includegraphics[width=0.50\textwidth]{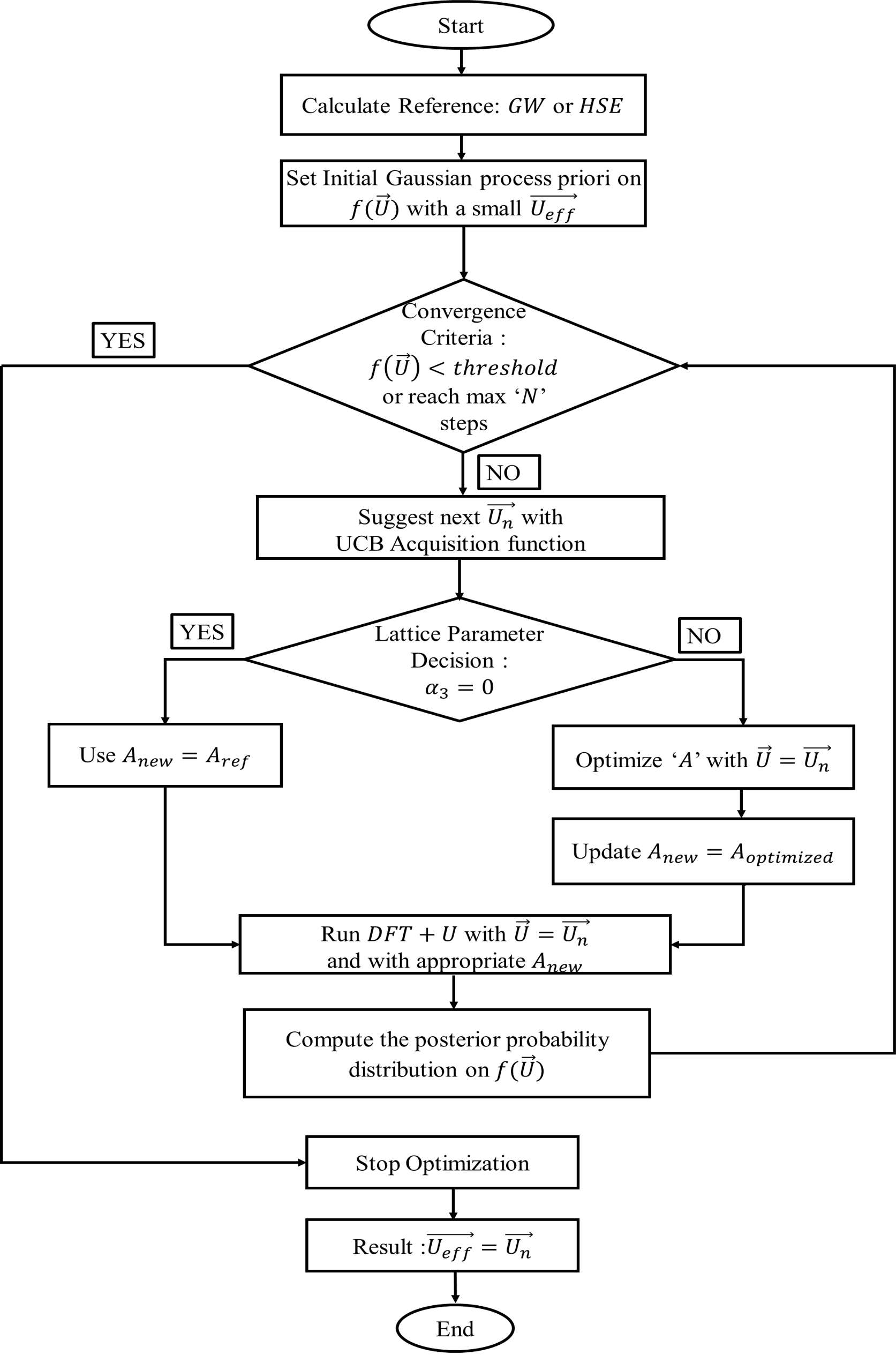}
    \caption{Flowchart of the BMach effective Hubbard $U$ parameter optimization algorithm.}
    \label{fig:flowchart}
\end{figure}

The flowchart in Figure \ref{fig:flowchart} outlines the steps of the BMach algorithm, beginning with an initial guess of the $U_{eff}$ parameters. The algorithm then uses Bayesian optimization to iteratively refine these values by minimizing the objective function, which includes terms for band gaps, eigenvalues, and lattice parameters. The convergence of the optimization process is achieved when the predicted values closely match the reference values from more accurate methods or experimental data.

\subsection[short]{Computational Tools and Techniques}

\begin{table*}[!htb]
    \caption{Effective Hubbard $U_{eff}$ parameters obtained with BMach using different methods and exchange-correlation (XC) functionals, with spin-orbit coupling (SOC) effects included. The energy of the first conduction band (band gap) $E_g (\Gamma_6^c-\Gamma_8^v)$, the second conduction band $E_g' (\Gamma_7^c-\Gamma_8^v)$, the valence band spin-orbit splitting $\Delta_{SO} (\Gamma_8^v-\Gamma_7^v)$, and the second conduction band spin-orbit splitting $\Delta_{SO}' (\Gamma_8^c-\Gamma_7^c)$ are evaluated at the $\Gamma$ point. Experimental results (with temperatures in Kelvin in parentheses) are taken or calculated from \cite{madelung_group_2002}. For hybrid HSEsol calculations, $\mu$ is the screening parameter that separates short-range and long-range interactions, and $\alpha$ is the exchange fraction that admixes PBEsol with the Hartree-Fock (HF) exchange.}
    \label{tab:ueff_eg}
    \begin{ruledtabular}
        \begin{tabularx}{\textwidth}{@{} l l c c c c c c c @{}}
            \multirow[c]{3}{*}{\textbf{Material}} & \multirow[c]{3}{*}{\textbf{Method}}                & \multicolumn{3}{c}{\textbf{$U_{eff}$ (eV)}} & \multicolumn{4}{c}{\textbf{Energy (eV)}}                                                                            \\
            \cmidrule{3-5} \cmidrule{6-9}         &                                                    & \textbf{In-5p}                              & \textbf{Sb-5p}                           & \textbf{As-4p} & $E_g$       & $\Delta_{SO}$ & $E_g'$   & $\Delta_{SO}'$ \\
            \midrule
            \multirow[t]{2}{*}{InAs}              & ONCV, PBEsol+U (BMach-2D)                          & -1.62                                       &                                          & 3.86           & 0.413       & 0.37          & 4.20     & 0.48           \\
                                                  & ONCV, PBEsol+U (BMach-1D)                          & -                                           &                                          & 3.72           & 0.413       & 0.37          & 4.16     & 0.49           \\
                                                  & ONCV, PBEsol+U (LR)                                & 0.78                                        &                                          & 4.70           & 0.670       & 0.38          & 4.35     & 0.50           \\
                                                  & Experiment (temp. in K) \cite{madelung_group_2002} &                                             &                                          &                & 0.415 (4.2) & 0.38 (1.5)    & 4.39 (5) & 0.47           \\
        \end{tabularx}
    \end{ruledtabular}
\end{table*}

\subsubsection[short]{\emph{Ab initio} Calculations}
All \emph{ab initio} structure relaxations and electronic band structure calculations were performed within the Density Functional Theory (DFT) framework using a Plane-Wave (PW) basis set and pseudopotentials-based method, as implemented in the \QE distribution \cite{giannozzi_advanced_2017}.

Brillouin-zone integrations were executed using $\Gamma$-centred k-point meshes.
For self-consistent field (SCF) calculations to determine the ground state charge density, and equilibrium lattice constants, an $8 \times 8 \times 8$ k-point grid was employed.
Gaussian smearing with a smearing width of 0.005 Rydberg was used in SCF calculations.
For non-self-consistent field (NSCF) calculations, a denser $16 \times 16 \times 16$ k-point grid was utilized.

For a demonstrative use of BMach the band structures $E(k)$ for zincblende (ZB) indium arsenide (InAs) were computed on a discrete k-mesh along high-symmetry directions.
Fixed occupations were used to manage electron occupation in NSCF and band structure calculations.
Spin-orbit coupling (SOC) corrections were applied to the interpolated bands for each eigenenergy $\varepsilon_{nk}$ and subsequently interpolated using maximally localized Wannier functions (MLWFs), following the method described by Malone and Cohen \cite{malone_quasiparticle_2013}.

\subsubsection[short]{Hubbard $U$ Calculation with LR Theory}
In this study, to determine the effective Hubbard parameter ($U_{eff}$) within the DFT+U framework using linear response (LR) theory, we used the HP code \cite{timrov_hp_2022} for calculating Hubbard parameters.
This approach leverages density-functional perturbation theory (DFPT) \cite{timrov_hubbard_2018} to compute Hubbard parameters without resorting to computationally prohibitive supercells.
Instead, it employs unit cells with monochromatic perturbations, facilitating a more computationally efficient yet precise determination of the parameters.
The HP code's theoretical framework allows perturbations to be expressed in reciprocal space, enabling the reconstruction of the response to localized perturbations as a series of monochromatic perturbations within a primitive unit cell.
The effective Hubbard parameters are derived from the second derivatives of the total energy with respect to atomic occupations, computed from the response matrices obtained through the self-consistent solution of the static Sternheimer equation.

For our SCF calculations performed using QE, we utilized a $\Gamma$-centered k-point mesh of $6 \times 6 \times 6$ to determine the ground state charge density and electronic structure, with a plane-wave cutoff energy set at 80 Ry to ensure precise results.
Non-collinear spin-orbit coupling (SOC) was considered in all calculations to capture the relativistic effects important for the materials studied accurately. Subsequently, the HP calculations employed an nq-point mesh of $2 \times 2 \times 2$ to apply monochromatic perturbations and compute the response matrices.

\section[short]{RESULTS AND DISCUSSIONS}\label{sec:results_discussions}

\begin{figure*}[!htb]
    \centering
    \begin{subfigure}[b]{0.40\textwidth}
        \centering
        \includegraphics[width=\textwidth]{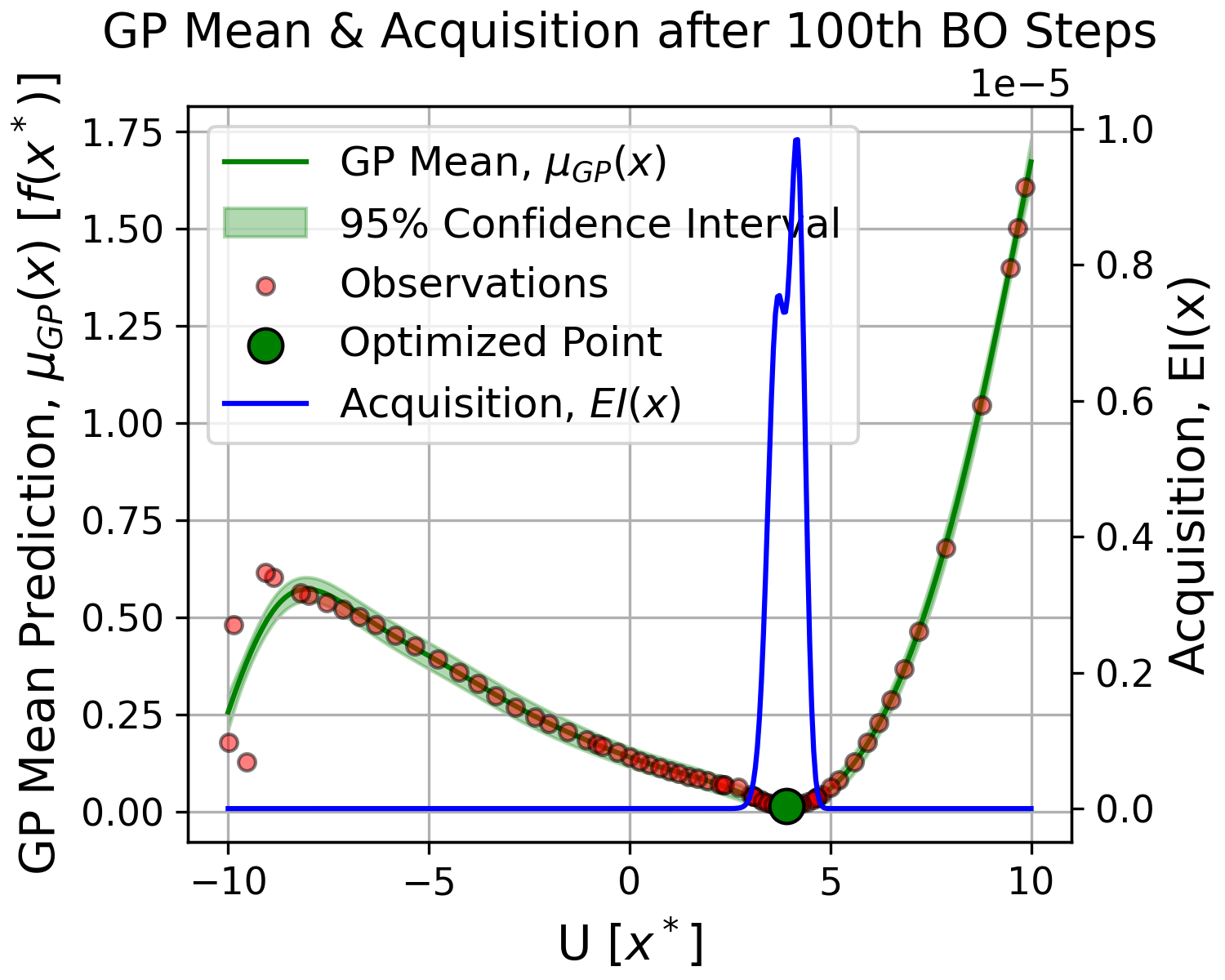}
        \caption{1D optimization results for InAs.}
        \label{fig:1d_optimization}
    \end{subfigure}
    \hspace{10mm}
    \begin{subfigure}[b]{0.35\textwidth}
        \centering
        \includegraphics[width=\textwidth]{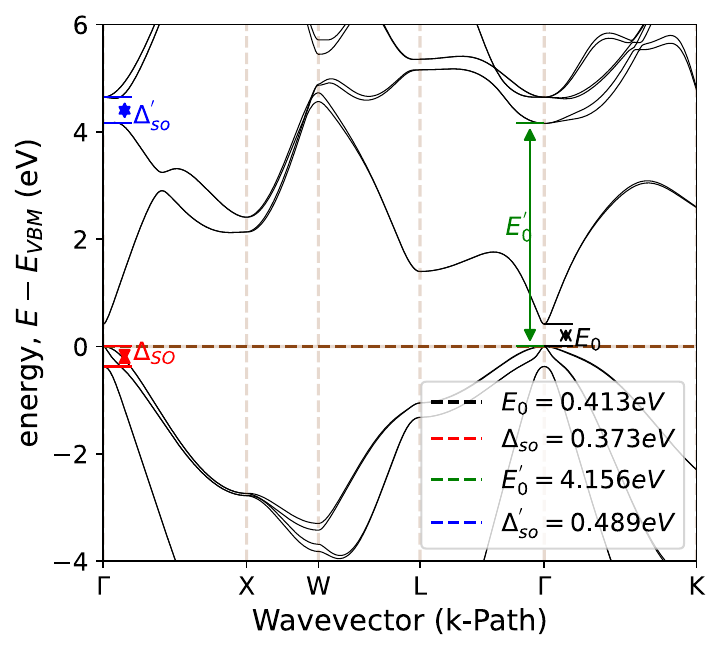}
        \caption{Band structure of InAs.}
        \label{fig:bs_InAs}
    \end{subfigure}
    \caption{(a) shows the optimization results for the 1D model of InAs, including the GP mean prediction, observations, optimized point, and acquisition function after 100 BO steps.
        (b) shows the band structure of InAs calculated using the DFT+U method with $U_{eff}^{As-4p}=3.72$ eV. The $U_{eff}$ values are calculated using BMach with $\alpha_3=0$ to minimize the objective function with the reference band structure. Experimental lattice parameter of 6.058~\AA\ for InAs is used for all calculations.}
\end{figure*}

\begin{figure*}[!htb]
    \centering
    \begin{subfigure}[b]{\textwidth}
        \centering
        \includegraphics[width=\textwidth]{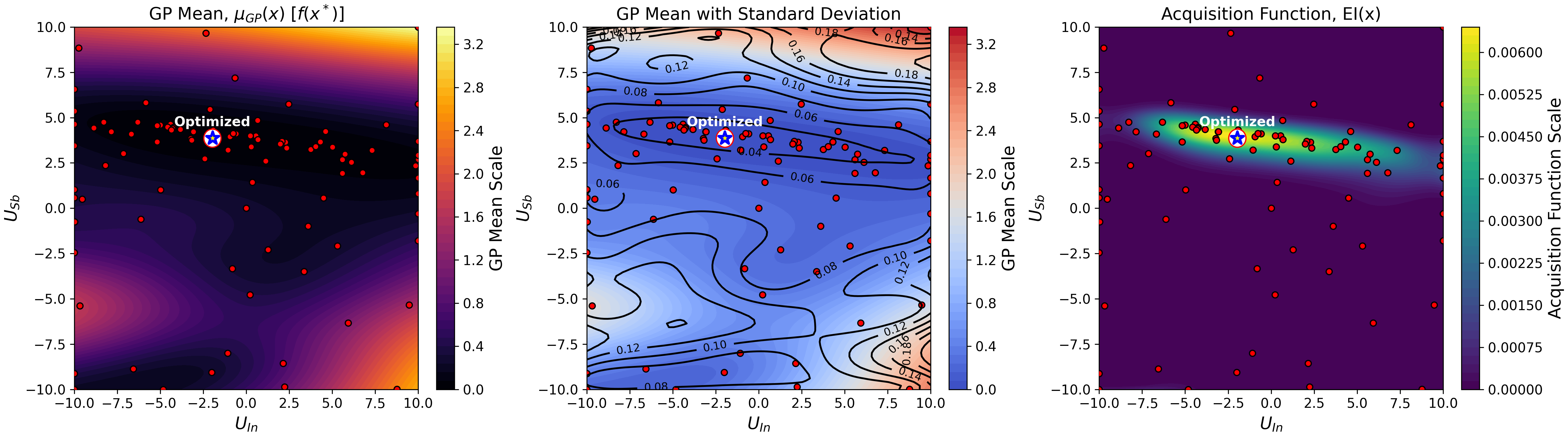}
    \end{subfigure}
    \caption{Optimization results for the 2D model of InAs after 100 BO steps. The figures show (a) GP mean prediction of the objective function $f(x^*)$, (b) GP mean with standard deviation showing the uncertainty in the prediction, and (c) acquisition function $EI(x)$ guiding the BO algorithm to the next sampling point. The red points represent observations, and the optimized point is highlighted with an asterisk, corresponding to the optimized 2D Hubbard $U_eff$ value of -1.62 eV for In-5p and 3.86 eV for As-4p.}
    \label{fig:2d_optimization}
\end{figure*}

The BMach algorithm, as detailed in the methods section (see \cref{subsec:bmach} and \cref{fig:flowchart}), was employed to optimize the effective Hubbard $U$ parameters for narrow band gap material InAs.

\subsection[short]{Narrow band gap semiconductor}
Narrow band gap materials are of significant interest due to their unique electronic properties, which make them suitable for various applications in optoelectronics and quantum computing. Among these materials, binary compound InAs is notable for its small band gap.

\subsubsection[short]{Indium Arsenide (InAs)}
The effective Hubbard $U$ parameters for InAs are optimized using BMach.
The band gaps ($E_g$) and spin-orbit splitting energies ($\Delta_{SO}$) are summarized in \cref{tab:ueff_eg}, and the corresponding band structures is shown in \cref{fig:bs_InAs}.

The LR results for InAs also reveal significant discrepancies.
The LR-calculated $E_g$ is 0.670 eV, which is higher than the BMach-optimized value of 0.413 eV but still misaligned with the experimental value of 0.415 eV.
These deviations indicate the limitations of the LR approach in capturing the electronic properties accurately.

\Cref{fig:1d_optimization} illustrates the optimization results for the 1D model of InAs.
The BMach algorithm efficiently converges to the optimal $U_{eff}$=3.72, providing accurate electronic and structural properties of InAs.
The GP mean values offer a smooth approximation of the objective function, while the acquisition functions guide the sampling of new points, balancing exploration and exploitation to achieve accurate and reliable predictions.

\Cref{fig:2d_optimization} illustrates the optimization results for the 2D model of InAs.
The optimized 2D Hubbard $U$ values are -1.62 eV for In-5p and 3.86 eV for As-4p, as indicated by the asterisk ($\ast$) in the figures, providing accurate electronic and structural properties for InAs.
The GP mean values offer a smooth approximation of the objective function, while the acquisition functions guide the sampling of new points, balancing exploration and exploitation to achieve accurate and reliable predictions.

Additionally, the margin of errors using BMach's Bayesian Optimization (BO) is very low, and the prediction of energies such as $E_g$, $\Delta_{SO}$, $E_g'$, and $\Delta_{SO}'$ are very close to experimental results, further confirming the reliability and accuracy of BMach.

\subsection[short]{Band Gap Optimization Results}
The developement of the effective Hubbard $U$ parameters' optimization process and its results using the BMach algorithm are illustrated in \cref{fig:1d_optimization} and \cref{fig:2d_optimization}.
These figures show the Gaussian process (GP) mean values, standard deviations, and acquisition functions during optimization for the 1D and 2D models of InSb and InAs, respectively.

The optimization plots highlight the efficiency of the BMach algorithm in converging to the optimal $U_{eff}$ values.
The GP mean values provide a smooth approximation of the objective function, while the acquisition functions guide the sampling of new points, balancing exploration and exploitation to achieve accurate and reliable predictions.

The results demonstrate that BMach effectively optimizes the Hubbard $U$ parameter for DFT+U calculations, yielding accurate electronic and structural properties for narrow-gap semiconductors like InSb and InAs.
The algorithm's application to alloy systems further underscores its versatility and predictive accuracy, making it a valuable tool for materials science research.

The incorporation of lattice parameters into the optimization process allows BMach to account for structural changes induced by variations in the Hubbard $U$ parameter.
This holistic approach leads to more accurate predictions of electronic properties and provides a comprehensive framework for modeling materials with strong electron correlation effects.

In comparison to traditional methods, the BMach algorithm offers several advantages.
It reduces the computational cost by efficiently exploring the parameter space and converging quickly to the optimal solution.
Additionally, BMach's ability to integrate multiple objectives, such as band gaps, eigenvalues, and lattice parameters, into a single optimization process enhances its robustness and applicability to a wide range of materials.

Overall, the BMach algorithm represents a significant advancement in the field of computational materials science.
Its ability to accurately determine the Hubbard $U$ parameter and predict electronic properties with high fidelity makes it a powerful tool for discovering and designing new materials.

\section[short]{CONCLUSIONS}\label{sec:conclusions}

In this study, we introduced the Bayesian Machine (BMach), a sophisticated algorithm designed to optimize the effective Hubbard $U$ parameter in DFT+U calculations by incorporating both electronic and structural properties.
Our approach leverages Bayesian optimization to iteratively refine $U_{eff}$, ensuring an accurate representation of material properties such as band gaps, eigenvalues, and lattice parameters.

The efficacy of BMach was demonstrated on narrow band gap semiconductor indium arsenide (InAs) for this moment.
Other materials are currently being tested under this framework, and the results will be updated in future versions of this manuscript.
The BMach-optimized $U_{eff}$ values yielded band gaps and spin-orbit splitting energies that align closely with experimental results, confirming the algorithm's reliability and accuracy.
This alignment underscores the importance of considering both electronic and lattice parameters in the optimization process, providing a more comprehensive model for materials with strong electron correlation effects.

The BMach algorithm offers significant advantages over traditional methods, including reduced computational cost and enhanced robustness in handling complex material simulations.
By integrating multiple objectives into a single optimization framework, BMach efficiently explores the parameter space and converges quickly to the optimal solution, making it a powerful tool for the discovery and design of high-throughput materials.

Future research can build on this work by applying BMach to a broader range of materials, including those with more complex electronic interactions and structural dynamics.
Additionally, further enhancements to the algorithm's acquisition functions and objective weighting strategies could further improve its predictive accuracy and computational efficiency.

Overall, BMach represents a significant advancement in computational materials science, paving the way for more accurate and efficient modeling of electronic properties in materials.
Its innovative approach and demonstrated success in optimizing the Hubbard $U$ parameter highlight its potential to drive forward the discovery and development of novel materials with tailored electronic and structural properties.

\vspace{8mm}
\section*{Data and code availability}
The data and code used to produce the results of this work will be made publicly available prior to publication.

\begin{acknowledgments}
    The computational work was conducted using the Ruche Supercomputing cluster at the Institute for Development and Resources in Scientific Computing (IDRIS), CNRS, Université Paris-Saclay.
    This research was supported by the French Ministry of Higher Education, Research and Innovationer (MESRI -- France) doctoral grant.
    I would like to acknowledge Dr. Zhenglu Li (formerly at Lawrence Berkeley National Laboratory, University of California Berkeley, and presently at the University of Southern California) for the insightful discussion that inspired the direction of this research.
    Finally, I thank my colleagues at C2N Lab, CNRS, Université Paris-Saclay for their support.
\end{acknowledgments}

\bibliography{bib/ref.bib}
\end{document}